# Optimizing Horticulture Luminescent Solar Concentrators via Enhanced Diffuse Emission Enabled by Micro-Cone Arrays

**Zhijie Xu[1], Martyna Michalska[2], and Ioannis Papakonstantinou [1] ***

[1] Photonic Innovations Lab, Department of Electronic and Electrical Engineering, University College London, London WC1E 7JE, UK.

[2] Manufacturing Futures Lab, Department of Mechanical Engineering, University College London, Queen Elizabeth Olympic Park, London, E20 3BS, UK.

*i.papakonstantinou@ucl.ac.uk

## Abstract

Optimizing the photon spectrum for photosynthesis concurrently with improving crop yields presents an efficient and sustainable pathway to alleviate global food shortages. Luminescent solar concentrators (LSCs), consisting of transparent host matrices doped with fluorophores, show excellent promise to achieve the desired spectral tailoring. However, conventional LSCs are predominantly engineered for photon concentration, which results in a limited outcoupling efficiency of converted photons. Here, we introduce a scheme to implement LSCs into Horticulture (HLSC) by enhancing light extraction. The symmetry of the device is disrupted by incorporating micro-cone arrays on the bottom surface to mitigate Total Internal Reflection (TIR). Both Monte Carlo ray tracing simulations and experimental results have verified that the greatest enhancements in converted light extraction, relative to planar LSCs, are achieved using micro-cone arrays (base width 50 µm, aspect ratio 1.2) with extruded and protruded profiles (85.15% and 66.55% improvement, respectively). Angularly resolved transmission measurements show that the HLSC device exhibits a broad angular radiation distribution. This characteristic indicates that the HLSC device emits diffuse light, which is conducive to optimal plant growth.

**Keywords:** luminescent solar concentrators, horticulture, outcoupling efficiency, micro-cones

## Introduction

Light is recognized as one of the paramount factors influencing plant growth[1–3]. Broadly, light's influence can be categorized into three key aspects: 1) light quantity, 2) light photoperiod, and 3) light quality[4]. Light quantity refers to the intensity of light received by plants within the Photosynthetically Active Radiation (PAR) spectrum, which spans from 400 to 700 nm and encompasses the wavelengths relevant to horticultural practices. Photoperiod is the amount of time plants receive light during a 24-hour period. Finally, light quality refers to the spectral distribution[5–7]. In brief, chlorophyll a and b, molecular pigments integral to photosynthesis of plants, exhibit strong absorption of blue and red wavelengths[8]. Conversely, ultraviolet and green can negatively impact crops by diminishing photosynthesis, decreasing shoot length, and reducing leaf absorption[9]. In other words, some parts of the PAR spectrum and particularly red are more beneficial for plant growth.

To accelerate plant growth in greenhouses, artificial lighting, such as from LEDs[10], has extensively been used. In principle though, natural light could also be manipulated to provide ideal lighting conditions for enhanced growth. Interestingly, the solar spectrum peaks in the green, whereas, as mentioned, the most useful wavelength in the PAR zone for plant growth is the red – see Figure 1b. An elegant solution to circumvent the misalignment between the two primary spectra, would be to shift the peak energy of AM1.5 from the green to the red. LSCs are widely employed spectral-shifting devices that can achieve this aim, due to the Stokes shift exhibited by the luminescent materials (or fluorophores) they contain[11,12]. As a result, several studies have recently appeared on the application of LSCs in horticulture[7,13,14]. LSCs not only facilitate spectral conversion but also confer the advantage of providing diffuse light for plant growth, given that fluorophores emit photons isotropically. Numerous prior investigations have substantiated the preference for diffuse light in cultivation[15], as unidirectional light is limited to certain portions of the plants, typically the leaves. Nonetheless, not only leaves but also stems and roots require solar radiation for optimal growth, a provision that is effectively met by diffuse light.

In a standard LSC device, fluorophores are hosted by a transparent dielectric material, such as polymer or glass[11,12,16–23]. As a result of the high refractive index contrast between the LSC and its surrounding (typically air), most converted light is concentrated within the device via TIR and it is guided towards its edges [24,25]. This is undesirable for our purposes however, since the objective here is to maximize the density of photons escaping and reaching the plants. In summary, apart from spectral conversion, light extraction is an equally critical component, but this is not effectively met by most HLSC technologies today.

Given the recent strides in fluorophore development over the past few decades, identifying appropriate green-to-red fluorophores is no longer a formidable challenge[26]. As a result, the primary gap that remains in achieving highly efficient HLSC lies in the implementation of novel light extraction techniques to enhance light quantity. In recent times, a multitude of techniques aimed at enhancing outcoupling efficiency have emerged within optical displays and OLEDs research[13,27–32]. The fundamental principle underlying these techniques revolves around the expansion of escape cones and/or the disruption of TIR cones[33]. The most straightforward approach to addressing this challenge involves reducing the refractive index of the host matrix, hence shrinking the TIR cone[32,34]. Another approach involves the use of nanostructures, such as gratings, metasurfaces, and photonic crystals, which overall help diffract waveguide modes out of the device[29,35,36]. A third option is the incorporation of micrometer relief structures on the surface of the LSC, which can enhance photon randomization, thereby creating more opportunities for light to enter the escape cone[37–39]. Among these options, microstructures are probably the most compatible solution with HLSC due to their potential scalability using current manufacturing methods, such as roll-to-roll hot-embossing[40]. Recently, a micro-lens array was applied to the top surface of an LSC device to extract about 30% of internally generated light from the bottom surface[7]. Alternatively, micro-cone arrays, widely used in numerous fields, from light outcoupling in optical backlights to light incoupling in solar cells can be used[41–43]. Inspired by these studies, micro-cone arrays are proposed here for HLSC purposes.

We employed Lumogen Red, an affordable and readily available organic dye for spectral conversion. It should be noted though, that our light extraction designs are widely applicable and not bound to any specific fluorophore, rendering them a universal solution for HLSC research. Supported by Monte Carlo ray tracing calculations, micro-cone arrays were first designed and optimized for effective light extraction. Several masters based on optimal designs were subsequently fabricated by 2-photon polymerization. Exact copies of the master surface relief structures (from now on called micro-cone extrusions) were created by a double inversion process, while a single inversion step created a negative copy (termed micro-cone protrusions). Even though protrusions exhibited a slightly lower outcoupling efficiency compared to extrusions, this design can potentially demonstrate superior robustness[44]. Micro-cone arrays for proof-of-concept (6×6 mm$^2$) were imprinted onto the bottom surface of PDMS-LSC (thickness of the sample is 3 mm), culminating in the realization of HLSC. The selection of PDMS as the host material, similarly to the choice of fluorescent materials, is primarily attributed to its availability and cost-effectiveness, rather than implying that PDMS is the exclusive material suitable for HLSC research. The enhancements in outcoupling efficiency for the converted red-light amounted to 85.15% and 66.55% for the best micro-cone extrusion and protrusion, respectively. Remarkably, angular distribution measurements demonstrated that light was emitted into a broad range of angles, a beneficial characteristic for promoting plant growth. Hence, our study paves the way for the practical implementation of HLSC technology.

**Results**

The schematic of HLSC realizing spectral conversion and improving photosynthesis is displayed in Figure 1a. Our proposed strategy involves employing fluorophores to capture green light and subsequently transform it into red light. Ideally, such a conversion should not compromise the photon counts of the solar radiation but rather optimize its spectral distribution. It is for this reason that fluorophores with high quantum efficiency are preferable. The absorption spectrum of Lumogen Red within HLSC was assessed using Ultraviolet–visible spectroscopy (UV-VIS, Shimadzu, UV-3600i Plus UV-VIS-NIR Spectrophotometer), revealing a peak around 560 nm, as shown in Figure 1b. Additionally, the emission spectrum was determined through time correlated single photon count spectroscopy (TCSPC, Edinburgh Instruments, FLS1000 Photoluminescence Spectrometer), indicating a peak approximately at 595 nm. Alongside a Stokes shift of 35 nm from green to red light, the majority of re-emitted photons falls within the 600 to 700 nm range. This range aligns remarkably well with the most effective spectral range for photosynthesis, as depicted by a shaded gold region in Figure 1b. To quantitatively illustrate this spectral conversion, we compared the proportion of energy in the 600-700 nm wavelength range before and after conversion within the entire PAR spectrum. This proportion increases from 33.6% in the AM1.5 spectrum to 46.1% in the converted spectrum. Furthermore, within the entire emission spectrum of Lumogen Red, the power in PAR constitutes a substantial proportion, accounting for up to 85.9%, thereby further enhancing the spectral distribution of light received by plants.

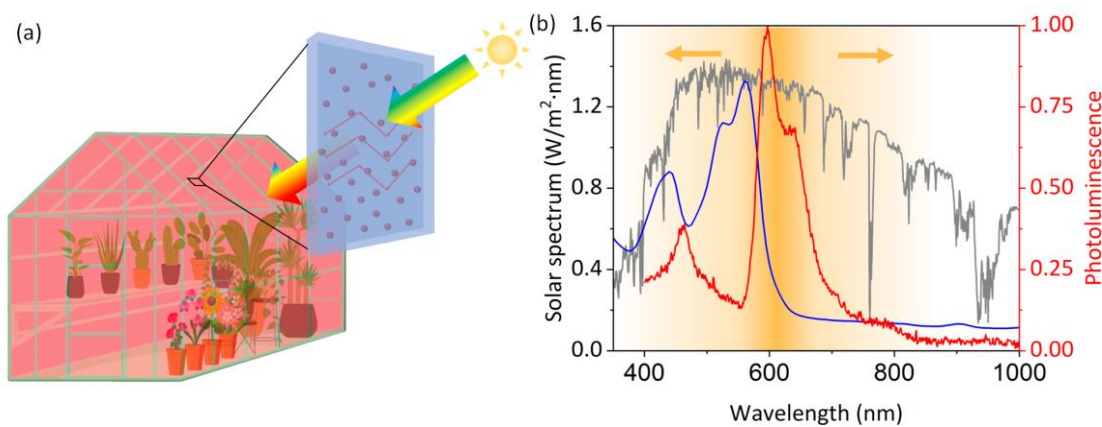

**Figure 1. a,** Illustration of a greenhouse covered by HLSC. **b,** Normalized absorption (blue line) and photoluminescence (red line) spectra of Lumogen Red in PDMS HLSC (dye concentration $1 \times 10^{-4}$ M). Photoluminescence was measured upon excitation at 350 nm. The solar spectrum is represented by the grey line, while the benefit for photosynthesis is indicated by the shaded gold region, diminishing with variations in wavelength (The golden arrows depict the trend of decay.).

While planar LSCs incorporating Lumogen Red can still attain efficient green-to-red conversion, their inherent ability to concentrate light constrains their applicability in horticulture. The possible photon fates in LSC devices are illustrated in Figure 2a[23]. In this context, we undertook a re-evaluation of the importance of photon fates, which differ from those typically reported in LSC research. This is because in conventional LSCs, the waveguide mode represents the preferred photon pathway, whereas the extracted photons from the bottom surface are the preferred photon fate in HLSC. As depicted in Figure 2a, if a photon is emitted outside the escape cone, it is trapped due to TIR. The escape cone forms two pairs of symmetrical cones for emission angles smaller than the critical angle. Absorption serves an ambivalent role. On one hand, high absorption effectively captures more light and transforms it into the desired range. On the other hand, it can result in undesirable high re-absorption rates.

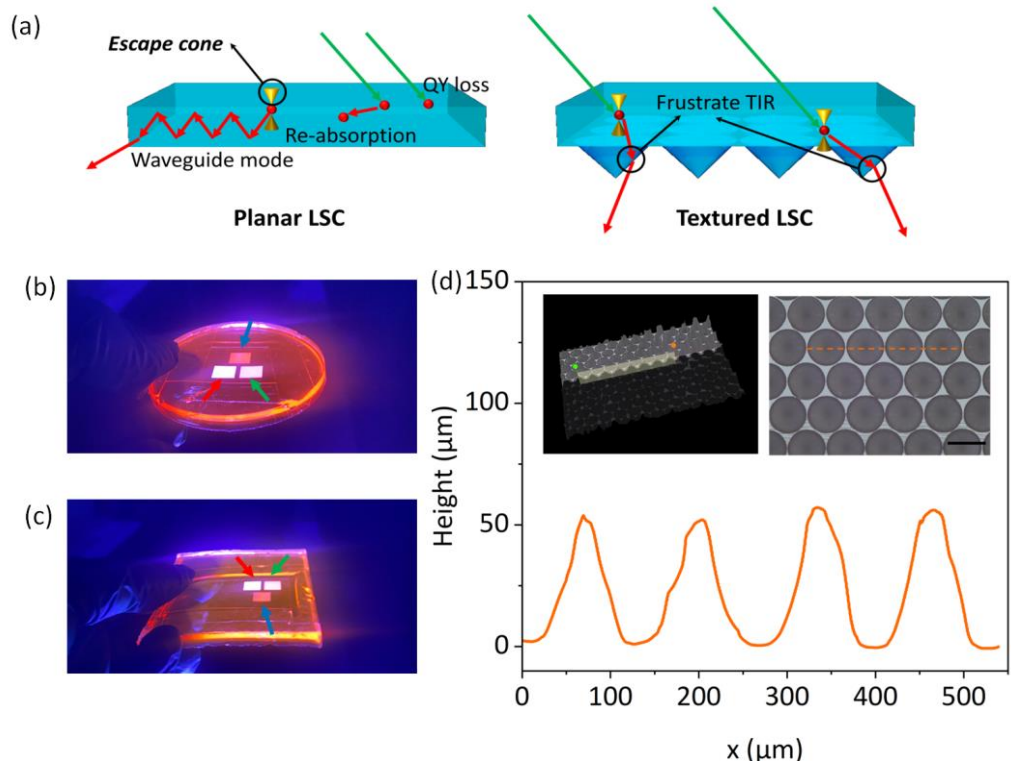

**Figure 2. a,** Theoretical explanations of photon pathways in LSC and HLSC enhancing light extraction using micro-cone arrays. **b-c,** Photographs illustrating the impact of **b** micro-cone extrusion and **c** protrusion on improving light extraction under ultraviolet illumination. **d,** Micro-cone array morphology examination utilizing a 3D optical microscope. The left inset presents a 3D optical micrograph, while the right inset provides a top-view perspective. Scale bar: 100 μm.

Based on the preceding discussion, it becomes imperative to establish definitions for photon fates within the context of HLSC, distinct from those applicable to conventional LSC devices. To be more specific, in the HLSC context, the internal quantum efficiency (IQE) is defined as the percentage of photons that escape from the bottom surface to the total number of absorbed photons. Concurrently, top loss and edge loss refer to the proportion of photons that escape from the top and edge surfaces respectively, in relation to the overall number of absorbed photons. The external quantum efficiency (EQE) on the other hand, characterizes the ratio of photons escaping from the bottom surface to the total number of incident photons. Likewise, external top loss and external edge loss signify the ratios of photons emanating from the top and edge surfaces relative to the total number of incident photons. Finally, unabsorbed loss designates the portion of photons traversing the device without undergoing absorption by fluorescent materials.

Building upon the aforementioned analysis of photon pathways in HLSC, the key challenge in achieving high-efficiency HLSC lies in mitigating the impact of the waveguide mode. Considering isotropic emission for the fluorophores, a substantial portion of photons—often exceeding 70%—tends to enter the waveguide mode for most common host materials, for which their refractive index is n~1.5[45]. Therefore, incorporating a light extraction technique is necessary to mitigate light trapping.

Extruding and protruding micro-cone arrays, like the ones shown in Figure 2a, are validated as effective microstructures for extracting light from high-index host matrices. This approach has the capability to disrupt the device's symmetry and promote light outcoupling. This is because even if a photon is originally emitted within the TIR cone, the reflectance angle would gradually change upon encountering the micro-cone array, until it falls into the escape cone. More details of photon paths are depicted in Figure S1 in supporting information.

We employed Monte Carlo ray tracing to identify the optimal structural parameters. Subsequent analysis determined that an HLSC device featuring a height-to-radius ratio (H/R) of 1.2 and a fluorophore concentration of $1\times10^{-4}$ M achieves the most favorable light extraction performance. The sample thickness is 3 mm, and the cone radius is fixed at 50 µm. The rationale for the selection of these parameters will be discussed in detail in subsequent sections. Simulated internal and external photon fates are depicted in Figure S2.

To fabricate HLSCs with micro-cone extrusions and protrusions, a multi-step fabrication approach was employed. Illustrated in Figure 3, periodic micro-cone arrays were first 3D-printed onto a polished silicon (Si) wafer by 2-photon polymerization (2PP, Photonic Professional GT, Nanoscribe). Following this step, the Si wafer with the micro-cone arrays served as a template to transfer the pattern into dye-doped PDMS via soft-lithography, ultimately producing protruding HLSC. Alternatively, nanoimprint lithography (NIL) was employed to first imprint protruding micro-cone arrays into an intermediate polymer stamp (IPS). Following a double inversion process using soft lithography in doped PDMS, an exact copy of the original extruding micro-cone arrays was created. Through the manipulation of the H/R ratio and fluorophore concentration, it becomes possible to tailor HLSCs to exhibit varying light extraction performances. For comparison, micro-cone arrays featuring three distinct H/R ratios (0.4, 1.2, and 2.0) and three different concentrations ($1\times10^{-4}$ M, $3\times10^{-5}$ M, and $1\times10^{-5}$ M) were fabricated.

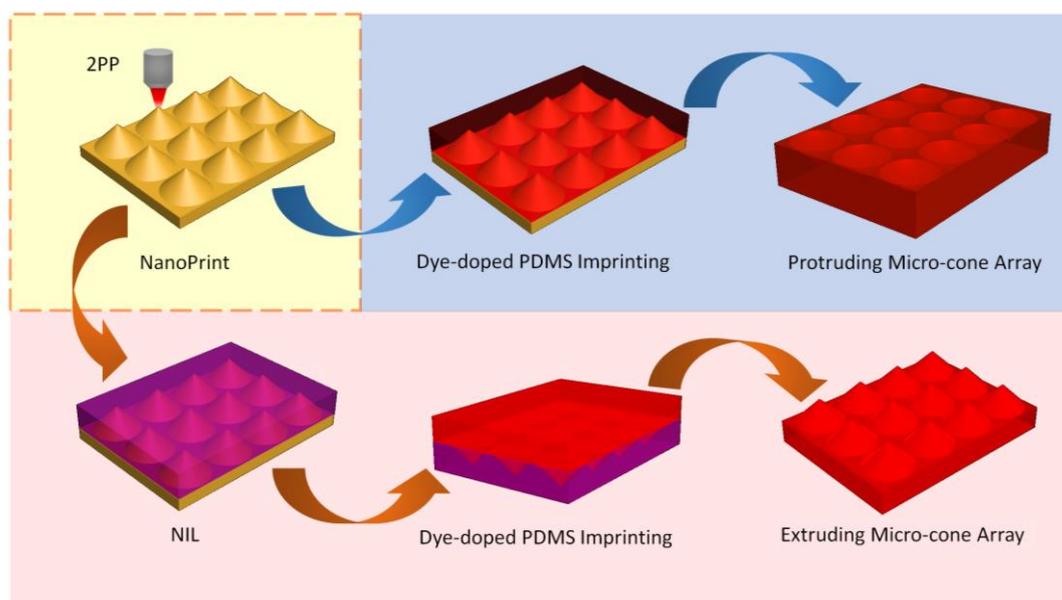

**Figure 3.** Illustrations depicting the fabrication process of HLCs with micro-cone extrusions and protrusions. The blue area and arrows indicate the fabrication process of protruding arrays, while the red area and arrows indicate the process of extruding arrays. The nanoprinted master for these two processes is marked within the yellow area surrounded by dashed line.

The surface morphology of the micro-cone arrays was assessed using a 3D optical microscopy system (Keyence, VHX 5000). A cross-section of four micro-cones, featuring a radius of 50 µm and a height of 60 µm (yielding an H/R ratio of 1.2), is illustrated in Figure 2d. This outcome underscores the successful realization of our structure design through nanoprinting, NIL, and PDMS soft-lithography processes. The left inset within Figure 2d provides a 3D profile of the micro-cone array from a 45° viewing angle. Furthermore, the right inset presents a top view of the micro-cone array, offering a glimpse of the cone base profile. Additional characterization of the structures is provided in the supporting information.

**Discussion**

Given that the fluorophores within HLSC are capable of absorbing incident photons from all directions, we utilize normally incident light to showcase the characteristics of HLSC. Additional Monte Carlo calculations have confirmed that the internal photon fates of HLSC under different incident angles remain nearly identical (see Figure S4). The outcoupling performance of the fabricated HLSCs was assessed by measuring the transmitted photon counts from the bottom surface. Visual representation of light extraction is provided in Figure 2b and 2c through photographs. The photographs clearly demonstrate that both micro-cone extrusions and protrusions exhibit significantly higher brightness compared to the surrounding planar LSC area. In the images, red arrows indicate the areas patterned by micro-cone arrays with an H/R ratio of 1.2, green arrows mark the areas with an H/R ratio of 2.0, and blue arrows designate the areas with an H/R ratio of 0.4. It is evident that there is a noticeable increase in brightness from the blue areas to the green areas and finally to the red areas, which is consistent with the quantitative results obtained in the subsequent analysis.

Photon counts were recorded using an integrating sphere (Labsphere, RTC-060-SF). To ensure fair comparison, micro-cone arrays with varying H/R values and a reference LSC were integrated into the same sample. All results have been normalized to the peak photon count. We integrated the photon counts between 550 and 700 nm for comparing outcoupling efficiency. The results, as illustrated in Figure 4a and 4c (solid lines), reveal substantial enhancements in outcoupling efficiency for both extruding and protruding structures, reaching 85.15% and 66.55%, respectively compared to planar

LSCs. In detail, 43.09% (extrusion) and 41.91% (protrusion) of the converted red photons can escape from the bottom surface (see Figure S2), percentages that are significantly higher than those achieved with micro-lens arrays (approximately 30%)[7].

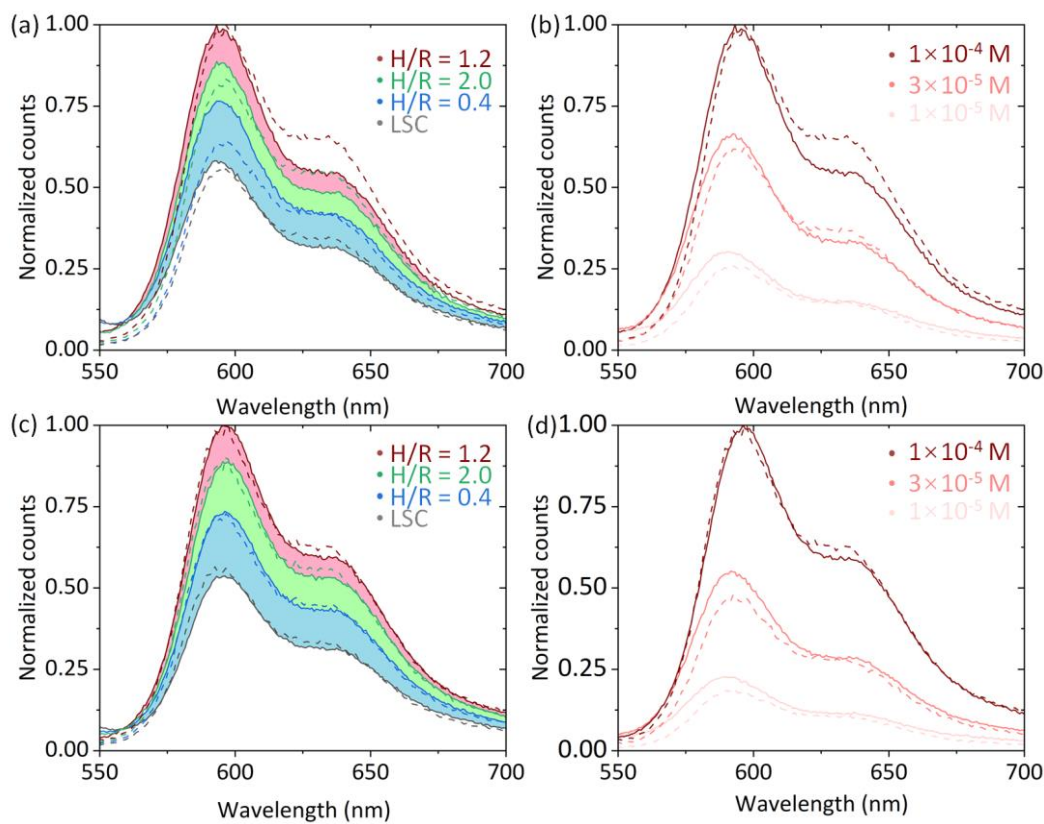

**Figure 4. Normalized counts of photons emitted from bottom surface, a,** Comparison between protruded micro-cone arrays with different H/R ratios and planar LSC. **b,** Comparison between different concentrations for protrusion. **c,** Comparison between extruded micro-cone arrays with different H/R ratios and planar LSC. **d,** Comparison between different concentrations for extrusion. Shades regions represent the outcoupling improvement. All solid lines correspond to experimental data, while dashed lines represent simulations.

Notably, all measured outcomes exhibit good agreement with Monte Carlo ray tracing predictions, as indicated by the dashed lines in Figure 4a and 4c. Across all H/R values, micro-cone arrays consistently display superior outcoupling efficiency when compared to planar LSCs. Of particular significance, the most substantial enhancement is observed at an H/R ratio of 1.2. This enhancement can be attributed to the combined effect of higher absorption and IQE. As outlined in Figure S2, the results from Monte

Carlo ray tracing affirm that an H/R ratio of 1.2 leads to the optimal absorption and IQE for micro-cone arrays. This notable absorption enhancement can be attributed to the structural capability of effectively recycling unabsorbed photons that reach the bottom surface. Additionally, based on Monte Carlo calculations, cones with high H/R ratios (≥ 1) exhibit a pronounced ability to extract rays. As shown in Figure S1, in the case of a high H/R ratio, even when the initial incident angle is greater than the critical angle, the micro-cone structure effectively reduces the incident angle through subsequent reflections. This ultimately results in the incident angle becoming smaller than the critical angle, allowing the photons to escape from the device. The results from Figure S2 indicate that a continuous increase in the H/R ratio leads to saturation in the improvement of IQE, reaching a critical threshold (H/R ~ 1).

Remarkably, a higher concentration is observed to yield superior performance, despite its potential for inducing increased re-absorption, as indicated in Figure 4b and 4d. This finding diverges somewhat from earlier research on LSCs, potentially simplifying future HLSC design by reducing the need for meticulous concentration optimization efforts. Furthermore, due to the pronounced re-absorption effect, the emission peak exhibits a notable red-shifting. Furthermore, when the concentration is $1 \times 10^{-4}$ M, the absorbance of HLSC reaches about 96% (excitation wavelength is 520 nm). As a result, increasing the concentration further may not be a more effective means of improving the performance of HLSC (see Figure S5). When the concentration is increased to $2 \times 10^{-4}$ M, both internal and external photon fates remain unchanged. However, if the concentration is raised to $3 \times 10^{-4}$ M or even higher, simulation results have shown that the portion of photons escaping from the top surface would exceed that of photons escaping from the bottom surface. As a result, further increasing the concentration yields diminishing returns in terms of enhancing the device's overall performance.

In a typical scenario, fluorophores emit light in an isotropic manner. Consequently,

a planar LSCs device would be akin to a Lambertian light source. To confirm this angular distribution, Bidirectional Transmittance Distribution Function (BTDF, IS-SA™, Radiant Zemax) measurement were conducted upon excitation at 350 nm, as demonstrated in Figure 5a. The measured BTDF is presented in polar coordinates (left inset) and agrees very well with Monte Carlo simulations (right inset). Both results are normalized to the peak irradiance for direct comparison. The cross-section along the dashed line on the measured BTDF is also plotted (black line) in Figure 5a and compared with an ideal Lambertian profile (red dashed line. The result confirms that the light output from the HLSC follows an almost perfect Lambertian profile), see the red area in Figure 5a. Note that the peak observed in the experimental cross-section arises from the part of the incident excitation that has not been absorbed by the HLSC (the green area in Figure 5 a and b). Since incidence is at a normal angle, a delta function appears at 0°. The reduction in the 0° peaks in Figure 5b, corresponding to the incident excitation from LSC to textured HLSC, serves as confirmation of the enhanced absorption of green light.

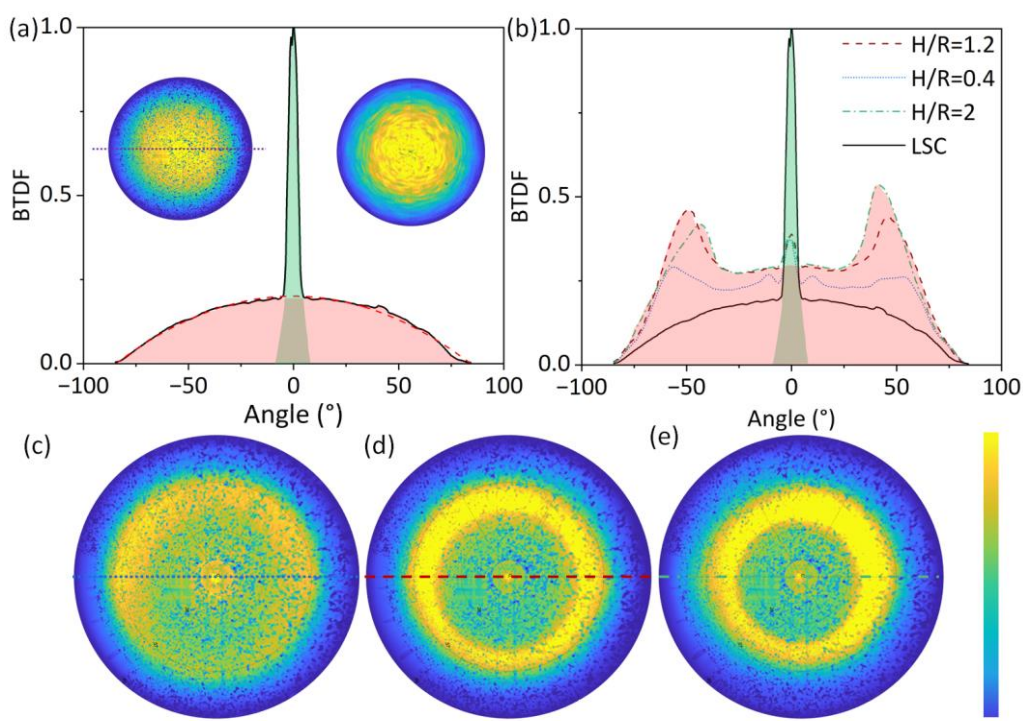

**Figure 5. a,** BTDF of planar LSC. Solid black line is the measured result from planar LSC, and dash red line is its Lambertian fitting. Insets show the BTDF of experiment and simulation. **b,** Cross sections of BTDF of different H/R ratios and planar LSC. Red areas in **a** and **b** represent emission, while the green areas depict the remaining excitation. **c-e,** BTDF of different H/R ratios.

The angular distribution of HLSC is also both measured and simulated. The comprehensive irradiance profiles are displayed in Figure 5c-e. Contrasting with the BTDF profile of the planar LSC, the most striking difference is the emergence of secondary rings at angles of 57°, 49°, and 42° for H/R 0.4, 1.2, and 2.0, respectively. This can be attributed to the geometry of the micro-cone arrays, which exhibit a preference for extracting light at angles nearly perpendicular to their surface. As H/R changes from 0.4 to 2.0, the cone angle decreases from 136° to 53°, resulting in a shift to the position of the secondary rings. The detailed ray tracing results in Figure S1 further explain the process. Furthermore, the overall irradiances stemming from the micro-cone arrays surpass those of the planar LSC across all emission angles. This outcome provides further confirmation of the enhancement in outcoupling efficiency through angular distribution. These distributions corroborate the capacity of HLSC to offer diffuse light conducive to plant growth.

As previously mentioned, aside from light quality, crop yield is also influenced by light quantity. This parameter is intricately tied to the direction of incident light. However, this is constantly changing throughout the day. For this reason, it is important to evaluate the performance of HLSC under different angles of incidence. As a realistic scenario, we used London to illustrate the correlation between incident angle and insolation—a parameter frequently employed to characterize the radiation reaching the Earth's surface. As depicted in Figure S7, if the incident angle falls above 45°, there is a marked decline in insolation. Consequently, our attention is directed towards incident angles below 50°, a range sufficient for effectively harnessing solar energy.

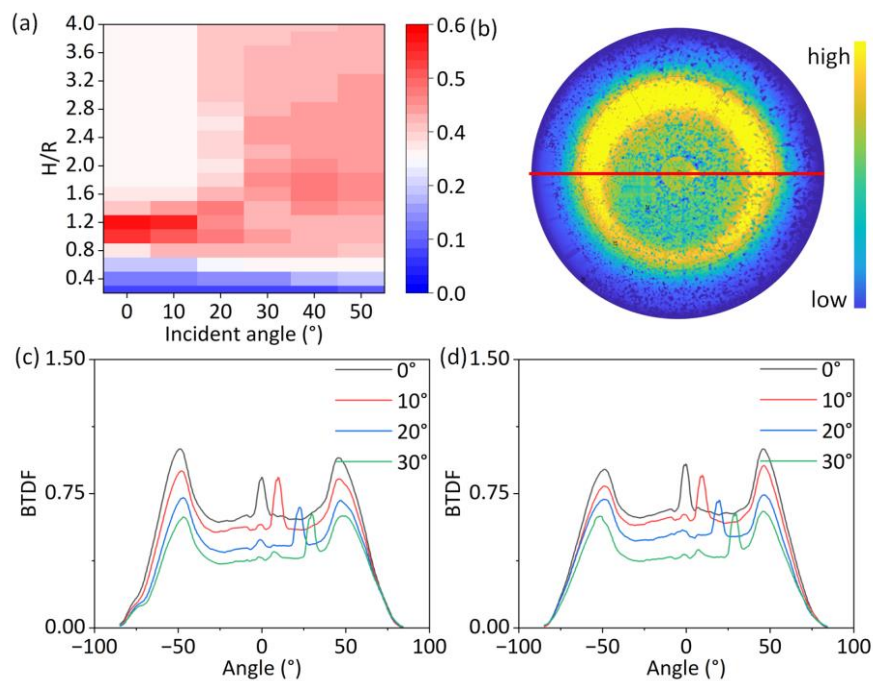

**Figure 6. a,** EQE enhancement for different H/R under distinct incident angles. **b,** BTDF in polar coordinates for micro-cone extrusions under an incident angle of 10°. **c-d,** Cross sections of BTDF under different incident angles for **c,** extruding and **d,** protruding micro-cone arrays.

The enhancement of EQE, in comparison to a planar LSC is presented in Figure 6a. Results reveal that HLSCs attains the highest EOE improvement, particularly when the H/R ratio is 1.2, across the majority of incident angles ranging from 0° to 50°. This finding reinforces the selection of this parameter as the optimal choice for incident angles. Furthermore, BTDF distributions are also calculated for varying incident angles, in Figure 6. The profiles of irradiance resemble those observed under normal incident conditions (see Figure S8), thereby continuing to provide diffuse light conducive to plant growth.

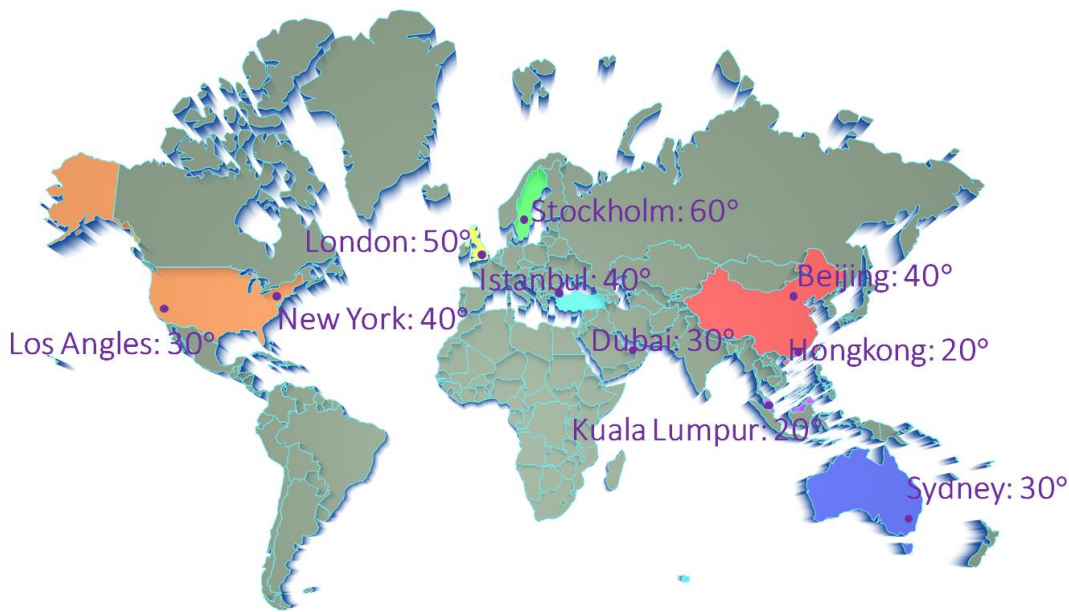

**Figure 7.** Recommended roof slopes of HLSC in different areas of world to achieve best photosynthesis efficiency.

To ensure that HLSC predominantly captures light from directions below 45°, we offer our recommendations for greenhouse roof slopes in various regions worldwide, as illustrated in Figure 7. As indicated by equations (s1-s8), insolation levels in various geographical regions are predominantly influenced by their respective latitudes. Consequently, the optimal roof slope varies in response to these latitude differences. This relationship is depicted in Figure 7, which illustrates that cities sharing a similar latitude tend to exhibit similar recommended roof slopes. For instance, regions located at lower latitudes, such as Kuala Lumpur, typically have relatively shallow roof slopes. In contrast, cities situated at higher latitudes, such as Stockholm, are expected to require steeper roof slopes to maximize solar exposure.

In summary, through the utilization of micro-cone array patterned HLSC, we have showcased a significant improvement in the outcoupling efficiency of converted red light. The incorporation of multi-step manufacturing techniques involving nanoprint and NIL has been detailed for the fabrication of PDMS-based HLSC devices. The collective evidence from both experimental and simulation results attests to the enhanced light extraction performance when compared to planar LSCs. Additionally, the measured angular distributions of HLSC devices underscore their capability to provide diffuse light, contributing to the optimal growth of plants.

## Materials and Methods

### Nanoprint and NIL

Micro-cone arrays with various H/R ratios are fabricated on a meticulously polished silicon wafer ($25 \times 25 \times 0.725$ mm$^3$). The wafer undergoes a thorough cleaning process using IPA and acetone solvents. Subsequently, it is placed within a vacuum chamber for a plasma treatment lasting 60 seconds. The formation of micro-cone arrays on the silicon wafer is achieved through two-photon polymerization (2PP) using a ×10 objective lens and a highly viscous liquid negative-tone resin known as IP-Q. Both the silicon wafer and IP-Q resin were procured from NanoScribe. Recognizing that 2PP is a considerably time-intensive procedure, modifications were made to the micro-cone arrays, transforming them into scaffolding structures. This adjustment allows for the retention of the micro-cone profiles for subsequent imprinting while significantly reducing the polymerization time.

Intermediate polymer stamp (IPS) is employed to replicate the micro-cone arrays obtained via Nanoprint through a NIL process. The sample is introduced into the nanoimprint facility (EITRE3, Obducat), where the imprinting process is conducted. The temperature is set to 170°C, and a pressure of 20 Bar is applied for 60 seconds to execute the imprinting. Subsequently, the temperature is reduced to 100°C for 20 seconds. Finally, the temperature is further lowered to 70°C, and the pressure is released to atmospheric levels. The entire NIL process takes approximately 10 minutes, significantly shorter than the nanoprint process, which typically requires around 6 hours to cover a $6 \times 6$ mm$^2$ area. Additionally, the use of IPS offers advantages in terms of durability and cost-effectiveness, making it well-suited for large-scale manufacturing.

**HLSC fabrication**

Lumogen Red (Sun Chemical Limited) is dissolved in ethyl acetate, which is compatible with PDMS. The dye solution is then added to the elastomer at the desired concentration and mixed using a magnetic stirrer. Subsequently, the mixture is placed in an ultrasonic bath at a temperature of 6°C for one hour. Following this, the curing agent (Sylgard 184, Dow Corning) is introduced into the mixture in a weight ratio of 1:10 compared to the elastomer. After degassing, the mixture is poured into a glass mold, the bottom surface of which is covered with either the nanoprint or NIL sample. The mold is then placed on a hot plate set at a temperature of 60°C for 2 hours to produce the final HLSC.

**Acknowledgements**

We are grateful to the Chinese Scholarship Council (CSC 202009110161) for the award of a PhD studentship and UCL Faculty of Engineering for the award of a Dean's prize. We thank Dr. Barry Reid and UCL Centre for Nature Inspired Engineering (CNIE) for help with 2-photon polymerization.

**Author Contributions**

Z.X. carried out structure fabrication and simulation. I.P. conceived the project. Z.X. conducted the optical measurements. M.M. helped with fabrication and characterization. Z.X. drafted the manuscript with help from I.P.

**Conflict of interest**

The authors declare no competing interests.

**Supplementary information**

Supplementary materials are available at the online version.